\documentclass[12pt,draftclsnofoot,journal,onecolumn]{IEEEtran}

\usepackage{cite}
\usepackage{graphicx}
\usepackage{amsmath}
\usepackage{amsfonts}
\usepackage{array}
\usepackage{subfigure}
\newtheorem{lemma}{Lemma}
\newtheorem{as}{Assumption}
\newtheorem{mdef}{Definition}

\begin{document}

\title{Networked MIMO with Clustered Linear Precoding}

\author{Jun~Zhang, Runhua~Chen, Jeffrey~G.~Andrews,
Arunabha Ghosh, and Robert~W.~Heath, Jr.
\thanks{J. Zhang, J. G. Andrews, and R. W. Heath are with the Wireless Networking and Communications
Group, Department of Electrical and Computer Engineering, The
University of Texas at Austin. Email: \{jzhang2, jandrews,
rheath\}@ece.utexas.edu. R. Chen is with the DSPS R\&D Center, Texas
Instruments Incorporated, Dallas. Email: r-chen@ti.com. A. Ghosh is
with AT\&T Labs. Email: ghosh@labs.att.com.}}

\maketitle

\begin{abstract}
A clustered base transceiver station (BTS) coordination strategy is
proposed for a large cellular MIMO network, which includes full
intra-cluster coordination--to enhance the sum rate--and limited
inter-cluster coordination--to reduce interference for the cluster
edge users. Multi-cell block diagonalization is used to coordinate
the transmissions across multiple BTSs in the same cluster. To
satisfy per-BTS power constraints, three combined precoder and power
allocation algorithms are proposed with different performance and
complexity tradeoffs. For inter-cluster coordination, the
coordination area is chosen to balance fairness for edge users and
the achievable sum rate. It is shown that a small cluster size
(about $7$ cells) is sufficient to obtain most of the sum rate
benefits from clustered coordination while greatly relieving channel
feedback requirement. Simulations show that the proposed
coordination strategy efficiently reduces interference and provides
a considerable sum rate gain for cellular MIMO networks.
\end{abstract}

\begin{keywords}
MIMO systems, cellular technology, resource allocation and
interference management, base station coordination.
\end{keywords}

\section{Introduction}
Multi-antenna transmission and reception (known as MIMO) is a key
technique for improving the throughput of future wireless broadband
systems. For a point-to-point link with multiple antennas at both
the transmitter and receiver, it has been shown that the capacity
grows linearly with the minimum number of transmit and receive
antennas, i.e. the number of spatial degrees of freedom\footnote{In
this paper, the definition of the number of spatial degrees of
freedom follows \cite{TseVis05}. It represents the dimension of the
transmitted signal as modified by the MIMO channel, and is equal to
the rank of the channel matrix when it has full rank. Therefore, for
a point-to-point link with $N_t$ transmit antennas and $N_r$ receive
antennas, it is $\min{(N_t, N_r)}$; for multiuser MIMO channels with
$K$ users, it is $\min{(N_t, KN_r)}$; for BTS coordination system
with $B$ BTSs, it is $\min{(BN_t, KN_r)}$.} \cite{Telatar99,
Foschini98}. Due to space constraints, however, mobile terminals
can only have a small number of antennas, normally one or two, which
bounds the capacity gain promised by MIMO. Multi-user MIMO
(MU-MIMO), where a BTS communicates with multiple mobile users
simultaneously, provides an opportunity to boost the sum capacity
through joint precoding (downlink) or joint decoding (uplink) even
when each user has only one antenna \cite{Goldsmith03}. For MU-MIMO
with a large number of mobile users, however, the sum capacity of
both the uplink and downlink is restricted by the number of antennas
at the BTS, as it determines the number of spatial degrees of
freedom.

Although theoretically attractive, deploying MIMO in a commercial
cellular system is fundamentally different as the transmission in
each cell acts as interference to other cells, and the entire
network is essentially interference-limited. While the problem of
interference is inherent to cellular systems, its effect on MIMO is
more significant because each neighboring BTS antenna element can
act as a unique interfering source, thereby making it difficult for
the mobile to estimate and suppress them. With $N_r$ receive
antennas, each mobile can only cancel/decode up to $N_r$ different
sources using linear techniques \cite{Paulraj03}. Furthermore,
interference is more severe for the downlink because complicated
interference suppression techniques are not practical for mobile
terminals, which need to be power-efficient and compact.
Coordination between users is usually not allowed. The capacity
gains promised by MIMO techniques have been shown to degrade
severely in the multi-cell environment
\cite{Catreux00,Blum03,Dai04}. Conventional approaches to mitigate
multi-cell interference, such as static frequency reuse, sectoring,
and spread spectrum, are not efficient for MIMO networks as each has
important drawbacks \cite{AndChoHea06}. The difficulty in combating
interference for MIMO is essentially due to the limitation of
spatial degrees of freedom, most of which are used to suppress the
spatial interference introduced by spatial multiplexing at the cell
site while few are left to suppress other-cell interference.

Thanks to the fast improvement of processing capability at BTSs and
the increase of the backhaul capacity, coordinated multi-cell MIMO
communications with cooperative processing among BTSs have drawn
significant amount of interest in recent years. The conventional
MIMO network with single-cell processing forms a MIMO interference
channel, whose spatial degrees of freedom are determined by the
number of transmit antennas at each BTS \cite{JafFak07IT}. With full
coordination across $B$ BTSs and a large number of mobile users, the
coordination system forms a virtual MIMO broadcast channel, which
increases the spatial degrees of freedom by $B$ times. Similar to
the transition from single-user MIMO to MU-MIMO, such cooperation
across multiple BTSs can provide great advantages over single-BTS
processing \cite{Shamai04,ShamaiVTC01,Zhang04,Karakayali06}. This
paper proposes a BTS coordination strategy with clustered linear
precoding for the downlink of cellular MU-MIMO systems, which
efficiently reduces interference provides a great sum rate gain by
exploiting the expanded spatial degrees of freedom.

\subsection{Related Work}
\emph{Intercell scheduling}, where neighboring BTSs cooperatively
schedule their transmissions, is a practical strategy to reduce
interference, as each time slot only one BTS in each cluster is
transmitting and it only requires message change comparable to that
for handoff. In \cite{Choi06TDMA}, it was shown that one major
advantage of intercell scheduling compared with conventional
frequency reuse is the expanded multiuser diversity gain. The
interference reduction is at the expense of a transmission duty
cycle, however, and it does not make full use of the available
spatial degrees of freedom.

Recently, \emph{BTS coordination} has been proposed as an effective
technique to mitigate interference in the downlink of multi-cell
networks \cite{Shamai04}. By sharing information across BTSs and
designing downlink signals cooperatively, signals from other cells
may be used to assist the transmission instead of acting as
interference, and the available degrees of freedom are fully
utilized. In \cite{ShamaiVTC01}, BTS coordination with DPC was first
proposed with single-antenna transmitters and receivers in each
cell. BTS coordination in a downlink multi-cell MIMO network was
studied in \cite{Zhang04}, with a per-BTS power constraint and
various joint transmission schemes. The maximum achievable common
rate in a coordinated network, with zero-forcing (ZF) and DPC, was
studied in \cite{FosHua05CISS,Karakayali06}, which demonstrated a
significant gain over the conventional single BTS transmission. With
simplified network models, analytical results were derived for
multi-cell ZF beamforming in \cite{Somekh06} and for various
coordination strategies with grouped cell interior and edge users in
\cite{Jing07}. Studies considering practical issues such as
limited-capacity backhaul and asynchronous interference can be found
in \cite{MarFet07EW,MarFet07ICC,SanSo07ISIT,ZhaMeh08TWC}.

With BTSs coordinating for transmission, it forms an effective
MU-MIMO broadcast channel, for which DPC has been shown to be an
optimal precoding technique
\cite{Caire03,Vishwanath03,Vishwanath03Tse,Yu04Cioffi,WeiSte06IT}.
DPC, while theoretically optimal, is an information theoretic
concept that is difficult to implement in practice. A more practical
precoding technique for broadcast MIMO channels is \emph{block
diagonalization (BD)}
\cite{Choi04,Spencer04,Pan04,CheAnd04ICC,ShiKwa07}, which provides
each user an interference-free channel with properly designed linear
precoding matrices. In addition, it was shown in \cite{SheChe07TWC}
that BD can achieve a significant part of the ergodic sum capacity
of DPC. Therefore, we will apply BD in the multi-cell scenario as
the precoding technique for the proposed BTS coordination.

Most previous studies on BTS coordination assume a global
coordination which eliminates inter-cell interference completely.
However, in realistic cellular systems, issues such as the
complexity of joint processing across all the BTSs, the difficulty
in acquiring full CSI from all the mobiles at each BTS, and time and
phase synchronization requirements will make full coordination
extremely difficult, especially for a large network. Therefore, it
is of great interest to develop coordination schemes at a local
scale, to lower the system complexity and maintain the benefits of
BTS coordination. For the uplink, an overlapping coordination
cluster structure was proposed in \cite{Ven07PIMRC}, where each BTS
is at the center of a unique cluster and coordinated combining is
performed to suppress interference for the central BTS of each
cluster. With such an overlapping cluster structure, each user is in
the interior of a cluster and enjoys interference reduction, but the
cluster number is as large as the number of BTSs and it cannot be
easily extended to the downlink. In \cite{BocHua07PIMRC}, the
downlink coordination over a 3-cell cluster was investigated with
both ZF and DPC, but no inter-cluster coordination was considered.

\subsection{Contributions}
In this paper, we propose a clustered BTS coordination strategy for
the downlink of a large cellular MIMO network. With full
coordination within the same cluster, the available spatial degrees
of freedom are greatly increased, which are then used to reduce
inter-cluster interference and exploit the sum rate gain. This
strategy consists of a full intra-cluster coordination and a limited
inter-cluster coordination. The intra-cluster coordination results
in precoding across BTSs within the same cluster for MU-MIMO, while
the inter-cluster coordination is used to pre-cancel interference
for the users at the edge of neighboring clusters. In this way,
interferences for both cluster interior and cluster edge users are
efficiently mitigated. Meanwhile, the system complexity and CSI
requirements at the BTSs, which are on a cluster scale, are greatly
reduced compared to global coordination. As the main complexity is
at the BTSs, mobile users can enjoy a simple conventional receiver.
In addition, the universal frequency reuse is applied, and there is
no need for cell planning.

We apply multi-cell BD as the precoding technique for such
coordination. The precoder matrix design is modified from
conventional single-cell BD, for which we consider other-cluster
interference suppression. In contrast to the classical MIMO
broadcast channel, the BTS coordination system has a per-BTS power
constraint. As there is no closed-form solution for the power
allocation problem with such a power constraint, three different
power allocation algorithms are proposed. For inter-cluster
coordination, we show that there is a tradeoff between fairness and
sum rate while choosing the inter-cluster coordination area. It is
shown that a small cluster size (about $7$ cells) can achieve a
significant part of the sum rate gain provided by the clustered
coordination while greatly reducing channel information feedback
compared to global coordination. Simulations show that the proposed
coordination strategy improves the sum rate over conventional
systems and reduces the impact of interference for cluster-edge
users.

The BTS coordination considering two classes of users (edge and
interior) was also investigated in \cite{Jing07}, which derived
information-theoretic results based on a simplified Wyner-type
circular network model. In this paper, we consider a more practical
setting--a large tesselated 2-D network. We propose clustered
coordination based on low-complexity linear precoding, design
parameters for such coordination and demonstrate the achievable
performance with simulation. We have made some idealized assumptions
in this paper, such as perfect information about channel state and
interference. We demonstrate through simulation that the
coordination system is sensitive to imperfect channel knowledge. The
full investigation of these practical issues is, however, left to
future work.

\subsection{Organization}
The rest of the paper is organized as follows: In Section II, we
make necessary assumptions, and describe the proposed coordination
strategy and the received signal model. Section III presents the
precoding matrix design for multi-cell BD. The detail for
inter-cluster coordination and the associated parameter design are
stated in Section IV. Numerical results are presented in Section V
and conclusions are drawn in Section VI.

\section{System Model}
\subsection{Clustered MIMO Network Structure}
\begin{figure}[htb]
\centering
\includegraphics[clip=true,scale=0.8]{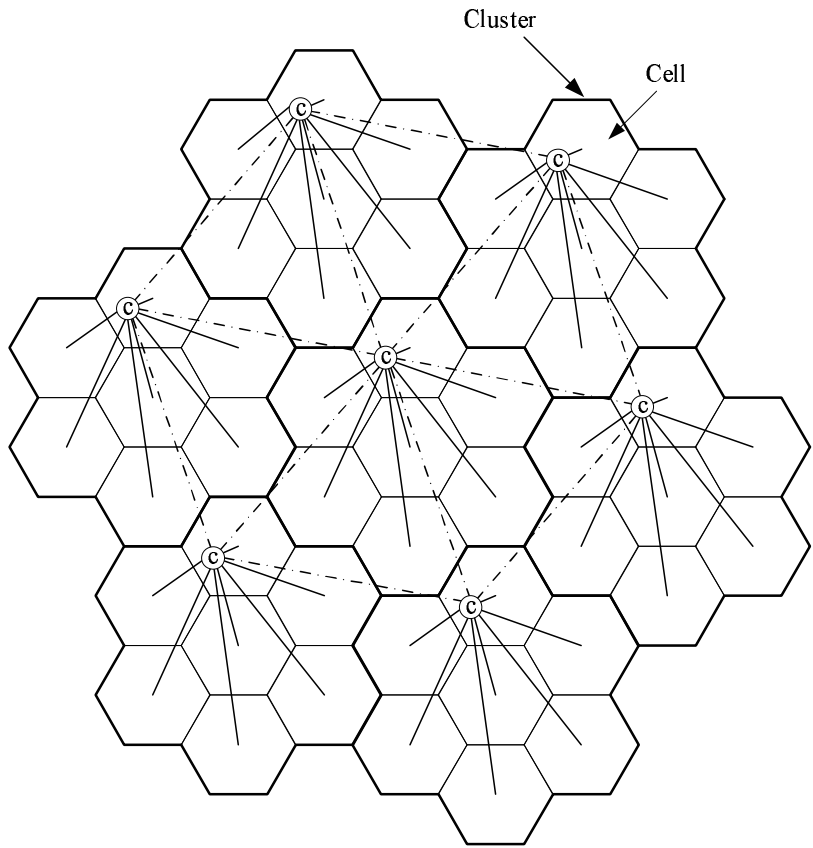}
\caption{An example of the clustered network, with $B=7$. Node ``C''
in each cluster is the virtual controller, which means full
coordination within each cluster. The dashed line between
controllers in neighboring clusters denotes the limited coordination
between these clusters.}\label{fig2_1}
\end{figure}

Consider a cellular MIMO network, where both BTSs and mobile users
have multiple antennas, $N_t$ and $N_r$, respectively. The system
parameters used in this paper are summarized in Table I. We consider
a large network, i.e. the number of cells in the network is very
large, so it is impractical to do coordination across all the BTSs.
We propose to divide the network into a number of disjoint clusters,
where each cluster contains a group of adjacent cells, as in Fig.
\ref{fig2_1}. With coordination among the BTSs within the same
cluster, we effectively increase the number of spatial degrees of
freedom, which will be used to suppress interference, including
inter-user and inter-cluster interference, and provide sum rate
gain.

We apply universal frequency reuse, so the users at the cluster edge
may suffer a high degree of interference from neighboring clusters.
To efficiently accommodate all the users, we group them into two
classes: cluster interior users and cluster edge users. A discussion
about user grouping will be given in Section IV. To do the proposed
clustered coordination, we make several assumptions.

\begin{as}
\emph{The BTSs within a cluster have perfect CSI of all the users in
this cluster, and perfect CSI of the edge users in the neighboring
clusters.}
\end{as}

For a time-division duplexing (TDD) system, the BTS can obtain the
downlink CSI through direct uplink channel estimation due to channel
reciprocity. For a frequency-division duplexing (FDD) system, the
downlink CSI can be obtained by feedback from mobile users, and
limited feedback for MU-MIMO is an ongoing topic
\cite{Jin06IT,RavJin07ICASSP,HuaHea07Tsp}, which we will not explore
in this paper and perfect CSI is assumed. The assumption of the
availability of CSI of the edge users is based on the fact that for
handoff such users have CSI of multiple neighboring clusters and can
feed back such information. The full CSI of the users in the same
cluster is for MU-MIMO precoding to cancel the inter-user
interference. The CSI of the edge users in the neighboring clusters
is for pre-canceling the inter-cluster interference for these users.

\begin{as}
\emph{The BTSs within the same cluster can fully share CSI and user
data. The BTSs in different clusters can exchange traffic
information, such as the number of active users and user locations.}
\end{as}

The capability of full coordination of the BTSs within the same
cluster enables doing MU-MIMO precoding across all the BTS antennas
in this cluster. The limited coordination between BTSs in different
clusters can be used for scheduling, e.g. the cluster with a large
number of active users may not do inter-cluster coordination for the
neighboring clusters.

\begin{as}
\emph{BTSs within the same cluster are perfectly synchronized in
time and phase, and different propagation delays from these BTSs to
mobile users in this cluster are compensated.}
\end{as}

This assumption is to ensure synchronous reception from the home
BTSs at mobile users. It is difficult to realize perfect
synchronization in practice, and the investigation of the impact of
asynchronous reception is out of the scope of this paper. Recently,
there has been some study on this subject \cite{ZhaMeh08TWC}.

From these assumptions, the system requirements for clustered
coordination are based on a cluster scale, which is much lower than
that for global coordination, especially in a large network.

\subsection{Coordination Strategy}
Based on the clustered structure and assumptions in the last
section, we propose a clustered coordination strategy, including
full intra-cluster coordination and limited inter-cluster
coordination. The transmission strategies for different user groups
are described as follows.
\paragraph*{Cluster interior users} BTSs within the same cluster work together as a ``super
BTS'' to serve the interior users in that cluster with MU-MIMO
precoding. In this way, there will be no intra-cluster interference,
i.e. inter-user interference, for these users. In addition, the
interior users are protected to a large degree from inter-cluster
interference by path loss.
\paragraph*{Cluster edge users} Multiple
neighboring clusters have channel information of edge users, and
they coordinate for the data transmission: one of these clusters is
selected to act as the home cluster to transmit data to such a user,
and other neighboring clusters will take this user into
consideration when designing precoding matrices. With
pre-cancelation of intra-cluster interference provided by the home
cluster and pre-cancelation of inter-cluster interference at other
neighboring clusters, there will be no interference for this edge
user from those clusters.

With such a coordination strategy, the interference for both cluster
interior and cluster edge users are efficiently mitigated.
Fractional frequency reuse (FFR) is another technique for
interference management where BTSs cooperatively schedule users in
different downlink bandwidths. However, FFR is a frequency-domain
interference management technique. The proposed coordination
strategy is a spatial domain technology that can be implemented with
a universal frequency reuse. For a highly-loaded system, FFR alone
cannot accommodate all the edge users. Networked MIMO offers another
opportunity to serve them.

\subsection{Received Signal Model}
Without loss of generality, we consider the cluster $c$. The
$N_r\times{1}$ received signal vector at the $k$th user in the
cluster $c$ is given as
\begin{equation}\label{equ2_1}
\mathbf{y}_k^{(c)}=\underbrace{\sum_{b=1}^{B}{\mathbf{H}_k^{(c,b)}{\mathbf{T}_k^{(c,b)}}\mathbf{x}_k^{(c)}}}_{\mbox{desired
signal}}+
\underbrace{\sum_{b=1}^{B}{\mathbf{H}_k^{(c,b)}\sum_{i=1,i\neq
k}^{K}{\mathbf{T}_i^{(c,b)}}\mathbf{x}_i^{(c)}}}_{\mbox{intra-cluster
interference}}
+\underbrace{\sum_{\hat{c}=1,\hat{c}\neq{c}}^{C}{\sum_{\hat{b}=1}^{B}{\mathbf{H}_k^{(\hat{c},\hat{b})}\sum_{j=1}^{K^{(\hat{c})}}{\mathbf{T}_j^{\hat{c},\hat{b}}\mathbf{x}_j^{(\hat{c})}}}}}_{\mbox{inter-cluster
interference}}+\mathbf{n}_k^{(c)}
\end{equation}
where
\begin{itemize}
\item{$\mathbf{x}_k^{(c)}$ is the $l_k\times 1$ transmitted vector for user $k$
in cluster $c$. Denote
$\bar{\mathbf{x}}^{(c)}=[\mathbf{x}_1^{(c)*}\quad\mathbf{x}_2^{(c)*}\quad\cdots$
$\mathbf{x}_K^{(c)*}]^*$, where $*$ denotes the conjugate transpose
of a matrix. The covariance matrix for $\bar{\mathbf{x}}^{(c)}$ is
denoted as
$\mathbf{Q}^{(c)}=\mathbb{E}[\bar{\mathbf{x}}^{(c)}\bar{\mathbf{x}}^{(c)*}]$.
}
\item{$\mathbf{H}_k^{(c,b)}$ is the $N_r\times N_t$ channel
matrix from BTS $b$ in cluster $c$ to user $k$.}
\item{$\mathbf{T}_k^{(c,b)}$ is the $N_t\times l_k$ precoding matrix
for user $k$ at the $b$th BTS in cluster $c$.}
\item{$\mathbf{n}_k^{(c)}$ is the additive white Gaussian nose at
user $k$ in cluster $c$, with zero mean and variance
$\mathbb{E}(\mathbf{n}_k^{(c)}\mathbf{n}_k^{(c)*})=\sigma_n^2\mathbf{I}_{N_r}$.}
\end{itemize}
Because the $B$ BTSs within this cluster coordinate to work as a
super BTS, the signal model can be written as
\begin{equation}\label{equ2_2}
\mathbf{y}_k^{(c)}=\mathbf{H}_k^{(c)}\sum_{i=1}^{K}{\mathbf{T}_i^{(c)}\mathbf{x}_i^{(c)}}+\mathbf{n}_k^{(c)}
+\sum_{\hat{c}= 1,\hat{c}\neq
c}^{C}{\mathbf{H}_k^{(\hat{c})}\sum_{j=1}^{K^{(\hat{c})}}{\mathbf{T}_j^{(\hat{c})}\mathbf{x}_j^{(\hat{c})}}}
\end{equation}
where
$\mathbf{H}_k^{(c)}=[\mathbf{H}_k^{(c,1)},\mathbf{H}_k^{(c,2)},\cdots,\mathbf{H}_k^{(c,B)}]$
is the $N_r\times{N_tB}$ aggregate channel transfer matrix from the
super BTS to user $k$, and
\begin{equation}\notag
\mathbf{T}_k^{(c)}=[\mathbf{T}_k^{(c,1)*}, \mathbf{T}_k^{(c,2)*},
\cdots, \mathbf{T}_k^{(c,B)*}]^*
\end{equation}
is the aggregate transmit precoder for user $k$ over all $B$ BTSs.
Unlike traditional downlink with co-located MIMO channels, the
channel gains from any two antennas at different BTSs are guaranteed
to be independent.

Denote $\mathbf{z}_k^{(c)}=\mathbf{n}_k^{(c)} +\sum_{\hat{c}=
1,\hat{c}\neq
c}^{C}{\mathbf{H}_k^{(\hat{c})}\sum_{j=1}^{K^{(\hat{c})}}{\mathbf{T}_j^{(\hat{c})}\mathbf{x}_j^{(\hat{c})}}}$
as the sum of the noise and interference from other clusters, the
covariance matrix of which is
\begin{align}
\mathbf{R}_k^{(c)}&=\sigma_n^2\mathbf{I}_{N_r}+\sum_{\hat{c}=
1,\hat{c}\neq
c}^{C}{\sum_{j=1}^{K^{(\hat{c})}}{\mathbf{H}_k^{(\hat{c})}\mathbf{T}_j^{(\hat{c})}\mathbb{E}[\mathbf{x}_j^{(\hat{c})}\mathbf{x}_j^{(\hat{c})*}]\mathbf{T}_j^{(\hat{c})*}\mathbf{H}_k^{(\hat{c})*}}}\notag\\
&=\sigma_n^2\mathbf{I}_{N_r}+\sum_{\hat{c}=1,\hat{c}\neq
c}^{C}{\sum_{j=1}^{K^{(\hat{c})}}{\mathbf{H}_k^{(\hat{c})}\mathbf{T}_j^{(\hat{c})}\mathbf{Q}_j^{(\hat{c})}\mathbf{T}_j^{(\hat{c})*}\mathbf{H}_k^{(\hat{c})*}}}.
\end{align}

\begin{as}
\emph{The interference plus noise covariance matrix is perfectly
known at the mobile users and BTSs in the same cluster.}
\end{as}

This covariance matrix can be estimated at mobile users by various
methods, including the usage of silent period of the desired signal
\cite{PadBat99Tcomm}, the usage of pilot signal \cite{KanBat98JSAC}
and blind estimation \cite{HonMad95IT} according to multiple access
strategies. After such estimation, each user will feed back it to
the BTS, which will be used to design precoding matrix.

\section{Clustered Multi-cell BD}
In the proposed coordination strategy, both cluster interior and
cluster edge users are served by multi-cell BD with pre-cancelation
at the ``super BTS''. BD is a linear precoding technique for
downlink MU-MIMO systems, and single-cell BD has been well studied
\cite{Choi04,Spencer04,Pan04,CheAnd04ICC,ShiKwa07}. A major
difference between multi-cell BD and single-cell BD is the power
constraint. While single-cell BD has a total power constraint (TPC),
each BTS in the cluster has its own power constraint, so multi-cell
BD has a per-BTS power constraint (PBPC). In this section, we will
design the clustered multi-cell BD, which can be separated into two
parts: the precoding matrix design and the power allocation design.
The design of the precoding matrix will consider other-cell
interference (OCI) and follow the algorithm proposed in
\cite{ShiKwa07}, which combines interference whitening at the
receiver and a statistical OCI-aware precoder at the transmitter to
reduce OCI and is shown to provide better sum rate performance than
conventional BD. For the power allocation, three different
algorithms will be proposed for PBPC.

\subsection{Precoding Matrix Design}
To suppress other-cell interference, we apply an $N_r\times{N_r}$
whitening filter $\mathbf{W}_k^{(c)}$ at the receiver for each user,
which is shown to be related with $\mathbf{R}_k^{(c)}$ as
\cite{ShiKwa07}
\begin{equation}\notag
\big[\mathbf{R}_k^{(c)}\big]^{-1}=\mathbf{W}_k^{(c)}\mathbf{W}_k^{(c)*}.
\end{equation}
With this whitening filter, the received signal for user $k$ after
post-processing is
\begin{equation}\label{eq_signal}
\mathbf{r}_k^{(c)}=\mathbf{W}_k^{(c)}\mathbf{H}_k^{(c)}\sum_{i=1}^{K}{\mathbf{T}_i^{(c)}\mathbf{x}_i^{(c)}}+\mathbf{W}_k^{(c)}\mathbf{z}_k^{(c)}
=\hat{\mathbf{H}}_k^{(c)}\sum_{i=1}^{K}{\mathbf{T}_i^{(c)}\mathbf{x}_i^{(c)}}+\hat{\mathbf{z}}_k^{(c)}
\end{equation}
where
$\hat{\mathbf{H}}_k^{(c)}=\mathbf{W}_k^{(c)}\mathbf{H}_k^{(c)}$ and
$\hat{\mathbf{z}}_k^{(c)}=\mathbf{W}_k^{(c)}\mathbf{z}_k^{(c)}$ are
equivalent channel matrix and noise vector.

Based on the equivalent signal model in \eqref{eq_signal}, we can
get the precoder for multi-cell BD. First, construct the aggregate
interference matrix for user $k$ in cluster $c$ as
\begin{equation}\label{MUIMatrix}
\tilde{\mathbf{H}}_k^{(c)}=\big[\hat{\mathbf{H}}_1^{(c)^*}\quad\cdots\quad\hat{\mathbf{H}}_{k-1}^{(c)^*}\quad\hat{\mathbf{H}}_{k+1}^{(c)^*}
\quad\cdots\quad\hat{\mathbf{H}}_K^{(c)^*}\big]^*.
\end{equation}
The principle idea of BD is to find the precoding matrix
$\mathbf{T}_k^{(c)}$ such that
$\tilde{\mathbf{H}}_k^{(c)}\mathbf{T}_k^{(c)}=\mathbf{0}$, which
means there is no inter-user interference. Thus $\mathbf{T}_k^{(c)}$
lies in the null space of $\tilde{\mathbf{H}}_k^{(c)}$. A sufficient
condition for the existence of a nonzero effective channel matrix
for user $k$, $\hat{\mathbf{H}}_k^{(c)}\mathbf{T}_k^{(c)}$, is that
at least one row of $\hat{\mathbf{H}}_k^{(c)}$ is linearly
independent of the rows of $\tilde{\mathbf{H}}_k^{(c)}$
\cite{HorJoh85}. This introduces the constraint that the number of
total transmit antennas ($BN_t$) is no smaller than the number of
total receive antennas ($KN_r$). Therefore, there is a constraint on
the total number of users that can be served simultaneously in each
cluster \cite{Spencer04,Pan04}, specified as follows \footnote{If
antenna selection or a decoding matrix is applied at the mobile
user, it is possible to support more users than this bound
\cite{Pan04, ChaHaz07}, which we will not consider in this paper.}:

\begin{lemma}[User constraint for multi-cell BD]\label{lemma1}\emph{For a clustered MIMO network with $B$ BTSs per
cluster, the maximum number of users that can be supported
simultaneously in each cluster by multi-cell BD is bounded by}
\begin{equation}\notag
K_{\textrm{max}}\leq \Big\lfloor\frac{BN_t}{N_r}\Big\rfloor.
\end{equation}
\emph{where $\lfloor{x}\rfloor$ is the maximum integer less than or
equal to $x$.}
\end{lemma}

Assuming $K\leq \Big\lfloor\frac{BN_t}{N_r}\Big\rfloor$, we describe
the precoding matrix design as follows. Let
$\tilde{l}_k=\mbox{rank}(\tilde{\mathbf{H}}_k^{(c)})$, and denote
the singular value decomposition (SVD) of
$\tilde{\mathbf{H}}_k^{(c)}$ as
\begin{equation}\notag
\tilde{\mathbf{H}}_k^{(c)}=\tilde{\mathbf{U}}_k^{(c)}\tilde{\mathbf{\Lambda}}_k^{(c)}
[\tilde{\mathbf{V}}_{k,1}^{(c)}\,\tilde{\mathbf{V}}_{k,0}^{(c)}]^*,
\end{equation}
where $\tilde{\mathbf{V}}_{k,1}^{(c)}$ contains the first
$\tilde{l}_k$ right singular vectors and
$\tilde{\mathbf{V}}_{k,0}^{(c)}$ contains the last
$B{N_t}-\tilde{l}_k$ right singular vectors. Therefore,
$\tilde{\mathbf{V}}_{k,0}^{(c)}$ forms a null space basis of
$\tilde{\mathbf{H}}_k^{(c)}$, from which we can get
$\mathbf{T}_k^{(c)}$. In this paper, we assume the number of spatial
streams for each user is $l_k=N_r$. If $l_k<N_r$ or there are extra
transmit antennas, additional optimization can be done by picking
the appropriate precoder subset \cite{Chen07} or doing coordinated
beamforming \cite{ChaHaz07}.

With the derived $\mathbf{T}_k^{(c)}$, the received signal becomes
\begin{equation}\notag
\mathbf{r}_k^{(c)}=\hat{\mathbf{H}}_k^{(c)}\mathbf{T}_k^{(c)}\mathbf{x}_k^{(c)}+\hat{\mathbf{z}}_k^{(c)}.
\end{equation}
Denote
$\bar{\mathbf{T}}_b^{(c)}=[\mathbf{T}_1^{(c,b)}\quad\mathbf{T}_2^{(c,b)}\quad\cdots\quad\mathbf{T}_K^{(c,b)}]$
as the submatrix associated with BTS $b$. Then the transmit power
constraint for each BTS is
\begin{equation}\notag
\mbox{Tr}(\bar{\mathbf{T}}_b^{(c)}\mathbf{Q}^{(c)}\bar{\mathbf{T}}_b^{(c)*})\leq
P.
\end{equation}
The achievable sum rate per cell for the clustered multi-cell BD is
then given by \cite{ShiKwa07}
\begin{equation}\label{equ3_4}
R_{CBD}=\max_{\mbox{Tr}(\bar{\mathbf{T}}_b^{(c)}\mathbf{Q}^{(c)}\bar{\mathbf{T}}_b^{(c)*})\leq
P}
\frac{1}{B}\sum_{k=1}^K{\log_2{\Big|\mathbf{I}_{N_r}+{\hat{\mathbf{H}}}_k^{(c)}{\mathbf{T}}_k^{(c)}\mathbf{Q}_k^{(c)}{\mathbf{T}}_k^{(c)*}{\hat{\mathbf{H}}}_k^{(c)*}\Big|}},
\end{equation}
where $\mathbf{Q}_k^{(c)}$ is the covariance matrix for
$\mathbf{x}_k^{(c)}$, and
$\big[\mathbf{Q}_1^{(c)*},\mathbf{Q}_2^{(c)*},\cdots;\mathbf{Q}_K^{(c)*}\big]^*=\mathbf{Q}^{(c)}$.

Denote the SVD of the effective channel
$\hat{{\mathbf{H}}}_k^{(c)}\mathbf{T}_k^{(c)}$ as
\begin{equation}\notag
\hat{\mathbf{H}}_k^{(c)}\mathbf{T}_k^{(c)}=\mathbf{U}_k^{(c)}
\left[\begin{array}{cc}\mathbf{\Lambda}_k^{(c)}&\mathbf{0}\\
\mathbf{0}&\mathbf{0}
\end{array}\right]
\mathbf{V}_{k}^{(c)},
\end{equation}
where
${\mathbf{\Lambda}}_k^{(c)}=\mbox{diag}(\lambda_{k,1},\cdots,\lambda_{k,{r}_k})$,
and $r_k=\mbox{rank}({\mathbf{H}}_k^{(c)}{\mathbf{T}}_k^{(c)})$. Let
$\mathbf{\Lambda}^{(c)}=\mbox{blockdiag}\{\mathbf{\Lambda}^{(c)}_1,\cdots,$
$\mathbf{\Lambda}^{(c)}_K\}$. Then the sum rate can be written as
\begin{equation}\label{CBD}
R_{CBD}=\max_{\mbox{Tr}(\bar{\mathbf{T}}_b^{(c)}\mathbf{Q}^{(c)}\bar{\mathbf{T}}_b^{(c)*})\leq
P}\frac{1}{B}{\log_2{\mid\mathbf{I}+{\mathbf{\Lambda}^{(c)}}\tilde{\mathbf{Q}}^{(c)}{\mathbf{\Lambda}^{(c)}}^{*}\mid}}
\end{equation}
where
$\tilde{\mathbf{Q}}^{(c)}={\mathbf{V}}^{(c)*}\mathbf{Q}^{(c)}{\mathbf{V}}^{(c)}$,
and
${\mathbf{V}}^{(c)}=\mbox{blockdiag}\{\mathbf{V}^{(c)}_1,\cdots,\mathbf{V}^{(c)}_K\}$.

\subsection{Power Allocation with PBPC} For the power allocation with PBPC, we propose one
optimal and two sub-optimal schemes: user scaling and scaled
water-filling. Both the optimal scheme and user scaling scheme are
convex optimization problems, and the scaled water-filling scheme is
modified from the conventional water-filling power allocation
algorithm.

\subsubsection{Optimal Power Allocation} The optimal power allocation
matrix to maximize \eqref{CBD} is a diagonal matrix
\cite{Telatar99}, denoted as
$\tilde{\mathbf{Q}}_{OPT}^{(c)}=\mbox{diag}(\gamma_{1,1},\gamma_{1,2},\cdots,\gamma_{1,l_1},\gamma_{2,1},\cdots,\gamma_{K,l_K})$.
The corresponding achievable sum rate is given as
\begin{equation}\notag
R_{CBD}=\max_{\mbox{Tr}(\bar{\mathbf{T}}_b^{(c)}\mathbf{Q}^{(c)}\bar{\mathbf{T}}_b^{(c)*})\leq
P}\frac{1}{B}\sum_{k=1}^{K}{\sum_{l=1}^{l_k}{\log\big(1+{\lambda^2_{k,l}}\gamma_{k,l}\big)}}.
\end{equation}

The power constraint can be rewritten as
\begin{equation}\notag
\sum_{k=1}^{K}{\sum_{l=1}^{l_k}{\|\mathbf{t}^{(c,b)}_{k,l}\|^2\gamma_{k,l}}}\leq{P},b=1,\cdots,B
\end{equation}
where $\mathbf{t}^{(c,b)}_{k,l}$ is the $l$th column of
$\mathbf{T}_k^{(c,b)}$.

Thus, the optimal power allocation problem with PBPC can be
formulated as
\begin{align}\label{Copt}
R_{CBD}=&\max_{\gamma_{i,j}}{\frac{1}{B}\sum_{k=1}^{K}{\sum_{l=1}^{l_k}{\log\big(1+{\lambda^2_{k,l}}\gamma_{k,l}\big)}}}\\
\mbox{subject to}&\left\{\begin{array}{l}
\sum_{k=1}^{K}{\sum_{l=1}^{l_k}{\|\mathbf{t}^{(c,b)}_{k,l}\|^2\gamma_{k,l}}}\leq{P},b=1,\cdots,B\notag\\
\gamma_{k,l}\geq 0, l=1,\cdots,l_k, k=1,\cdots,K.
\end{array}\right.
\end{align}
For this optimization problem, the power constraints for different
users are coupled. Similar problems with per-antenna power
constraints have been studied in \cite{Boccardi06,YuLan07Tsp}. To
the best of our knowledge, no efficient algorithm as water-filling
for power allocation problems with per-antenna or per-BTS power
constraints is available at this point. The objective function,
however, is concave and the constraint functions are linear, so this
is a convex optimization problem and can be solved numerically, e.g.
with the interior-point method \cite{Boyd04}. However, with a large
number of users, and multiple transmit and receive antennas, it is
quite complex to solve this optimization problem, and we propose two
sub-optimal schemes in the following sections.

\subsubsection{User Scaling (US)} One sub-optimal power allocation
scheme is user scaling, for which we weight the precoding matrix for
each user, by choosing
$\tilde{\mathbf{Q}}_{US}^{(c)}=\mbox{blockdiag}(\mu_1\mathbf{I}_{L_1},\mu_2\mathbf{I}_{L_2},\cdots,$
$\mu_K\mathbf{I}_{l_K})$, where $\mu_k$ is to scale the precoding
matrix of the $k$th user to meet the power constraint.

There are several reasons for doing this. First, with fewer weight
terms it reduces the complexity for solving the optimization problem
compared with the optimal scheme. Second, for each user an equal
power allocation only results in a negligible capacity loss compared
to the optimal water-filling, especially at high SINR, and with
shadowing the power allocation across users plays a more important
role than across streams of each user. Third, user scaling makes it
easy to adjust transmit power between different users, for example,
to meet a fixed rate constraint.

Denote ${\omega}_{k}^{(c,b)}=\|\mathbf{T}_k^{(c,b)}\|^2_F$. The
optimization problem for the user scaling scheme is
\begin{align}\label{Cus}
R_{US}=&\max_{\gamma_{i,j}}{\frac{1}{B}\sum_{k=1}^{K}{\sum_{l=1}^{l_k}{\log\big(1+{\lambda^2_{k,l}}\mu_{k}\big)}}}\\
\mbox{subject to}&\left\{\begin{array}{l}
\sum_{k=1}^{K}{\omega_{k}^{(c,b)}\mu_k}\leq{P},b=1,\cdots,B\notag\\
\mu_{k}\geq 0, k=1,\cdots,K.
\end{array}\right.
\end{align}
Again, this is a convex optimization problem.

\subsubsection{Scaled Water-Filling (SWF)} As it is difficult to get
an efficient algorithm to solve (\ref{Copt}) and (\ref{Cus}), we
propose another sub-optimal scheme based on the water-filling
algorithm.

First, consider a multi-cell BD system with TPC, whose sum rate is
given by
\begin{equation}\label{equ_TPC}
R_{TPC}=\max_{\mbox{Tr}({\mathbf{T}}^{(c)}{\mathbf{T}}^{(c)*})\leq
BP_{max}}\frac{1}{B}{\log_2{\mid\mathbf{I}+{\mathbf{\Lambda}^{(c)}}\tilde{\mathbf{Q}}_{TPC}^{(c)}{\mathbf{\Lambda}^{(c)}}^{*}\mid}}.
\end{equation}
The optimal power loading matrix
$\tilde{\mathbf{Q}}_{TPC}^{(c)}=\mathbf{\Sigma}^{(c)}$ is derived by
water-filling \cite{Spencer04}. To meet PBPC, we scale this matrix
and choose
$\tilde{\mathbf{Q}}_{SWF}^{(c)}=\mu\tilde{\mathbf{Q}}_{TPC}^{(c)}$.
The scaling factor $\mu\in (0,1)$ is given by
\begin{equation}\notag
\mu=\frac{P}{\max_{b=1,2,\cdots,B}{\mbox{Tr}({\bar{\mathbf{T}}}_b^{(c)}\tilde{\mathbf{Q}}_{TPC}^{(c)}{\bar{\mathbf{T}}}_b^{(c)*})}}
\end{equation}
Therefore, the sum rate per cell is given by
\begin{equation}\label{Cswf}
R_{SWF}=\frac{1}{B}\log_2{\mid\mathbf{I}+\mu\mathbf{\Lambda}^{(c)}\mathbf{\Sigma}^{(c)}\mathbf{\Lambda}^{(c)*}\mid}.
\end{equation}


\subsection{Scheduling Schemes} From \emph{Lemma \ref{lemma1}}, there is a
constraint on the maximum number of users a multi-cell BD system can
support simultaneously. Therefore, with a large number of users in
each cluster, it is necessary to schedule transmission for a subset
of users, according to some performance criterion. The sum rate
optimal scheduling algorithm is to exhaustively search over all the
possible user combinations and pick the user set which maximizes the
chosen performance metric, which is extremely complicated. We
propose to use a sum rate based sub-optimal user selection algorithm
inspired by \cite{Shen06}, which has low complexity and approaches
optimal performance.

Let $\mathcal{U}$ and $\mathcal{S}$ denote the sets of unselected
and selected users respectively, and $f_k$ denotes the performance
metric for user $k$. The proposed user selection algorithm is
described in Table II. This is a greedy algorithm. In each step, one
user is selected from the un-selected user set which adds the
maximum performance gain, and the process stops when no more user
can be added or the performance metric begins to decrease. We
consider two different kinds of scheduling, maximum sum rate (MSR)
and proportional fairness (PF), for different scenarios.

\section{Inter-cluster Coordination}
With the proposed coordination strategy, BTSs within a cluster serve
their interior users with multi-cell BD, while the neighboring
clusters coordinate with each other to serve edge users. It is
possible for multiple BTSs to transmit data to an edge user, but for
simplicity we consider that each user is served by one cluster. In
this section, we will describe inter-cluster coordination in detail,
and investigate two important system parameters: coordination
distance and cluster size.

\subsection{Inter-cluster Coordination with Multi-cell BD}
\begin{figure}[htb]
\centering
\includegraphics[clip=true,scale=0.6]{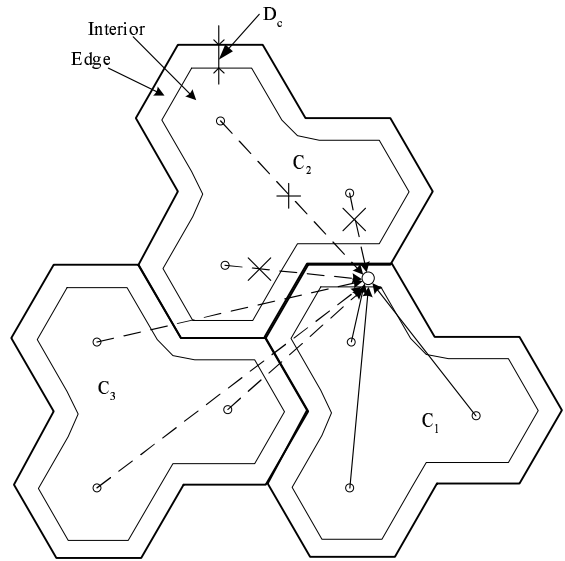}
\caption{An example of inter-cluster coordination, $B=3$. $C_1$ is
the home cluster, $C_2$ is the helper cluster, and $C_3$ is the
interferer cluster. Solid lines denote transmissions of information
signals and dotted lines are interference, and the cross on the
dotted line means that the interference is
pre-canceled.}\label{figICCdiagram}
\end{figure}

The main idea of inter-cluster coordination is to do interference
pre-cancelation at all the neighboring clusters for the active edge
user, and select one cluster to transmit information data to this
user. The precoding technique used in this paper for inter-cluster
coordination is multi-cell BD, the same as for intra-cluster
coordination. Each edge user selects a cluster based on the channel
state, denoted as the \emph{home cluster}, and feeds back this
decision, while the other neighboring clusters act as \emph{helpers}
for the data transmission. The remaining clusters are
\emph{interferer clusters}. Different kinds of clusters and
inter-cluster transmission are illustrated in Fig.
\ref{figICCdiagram}.

For the following discussion and simulation, we focus on a home
cluster and assume that when this cluster schedules an edge user,
the neighboring clusters of this edge user will always help. This
will happen if there are a small number of users in each cluster so
that there are spare degrees of freedom at neighboring clusters.
With a large number of users, joint scheduling across clusters is
required. While we leave the full investigation of such a scheduling
problem to future work, we propose a simple two-step approach:
first, each cluster does scheduling within its own cluster, and the
scheduled edge users inform the neighboring helper clusters; in the
second step, each cluster deals with the requests from edge users in
the neighboring clusters, and it selects to help some of these users
while drops some scheduled users of its own. After this scheduling
process, each cluster designs precoding matrices.

To the home cluster, there is no difference between the edge user
and interior users, and the BD precoding matrix is designed as in
Section III. For helper clusters, the precoding matrix design will
be different. Without loss of generality, we consider the precoding
matrix design at the helper cluster $c_1$ for the edge user $k_0$,
which is served by its home cluster $c_0$. Denote
\begin{equation}\notag
\bar{\mathbf{H}}_{k_1}^{(c_1)}=[\tilde{\mathbf{H}}_{k_1}^{(c_1)*}\quad
\hat{\mathbf{H}}_{k_0}^{(c_1)*}]^*,
\end{equation}
where $\tilde{\mathbf{H}}_{k_1}^{(c_1)}$ is the aggregate
interference matrix of user $k_1$ to all the other active users
\footnote{The active users in a cluster are the users currently
being served.} in the cluster $c_1$ as in \eqref{MUIMatrix}, and
$\hat{\mathbf{H}}_{k_0}^{(c_1)}=\mathbf{W}_{k_0}^{(c_0)}{\mathbf{H}}_{k_0}^{(c_1)}$
is the effective channel after whitening filter from the cluster
$c_1$ to the edge user $k_0$.

To pre-cancel the interference for both the edge user $k_0$ and
other active users in the cluster $c_1$, the precoding matrix
$\mathbf{T}_{k_1}^{(c_1)}$ should satisfy the condition
$\bar{\mathbf{H}}_{k_1}^{(c_1)}\mathbf{T}_{k_1}^{(c_1)}=\mathbf{0}$,
i.e. it should lie in the null space of
$\bar{\mathbf{H}}_{k_1}^{(c_1)}$, which can be designed with SVD of
$\bar{\mathbf{H}}_{k_1}^{(c_1)}$ in the same way as in Section III.
Similar to \emph{Lemma 1}, there is a constraint on the number of
users that can be supported simultaneously in the helper cluster,
stated as follows:
\begin{lemma}[User constraint for the helper cluster]\label{lemma2}
\emph{For a helper cluster with $B$ BTSs and $k_e$ edge users to
help, the maximum number of users that can be supported
simultaneously by multi-cell BD in this cluster is bounded by}
\begin{equation}\notag
K^h_{\textrm{max}}\leq \Big\lfloor\frac{BN_t}{N_r}\Big\rfloor-k_e.
\end{equation}
\end{lemma}

Therefore, to serve an edge user with inter-cluster coordination,
the total number of users the network can support will be reduced,
which induces a tradeoff between mitigating interference for edge
users and maximizing the total throughput. This makes the choice of
the inter-cluster coordination area important. Actually, the user
constraints in \emph{Lemma 1} and \emph{Lemma 2} are due to the
constraint on the total spatial degrees of freedom in each cluster,
determined by the cluster size and the number of transmit antennas
at each BTS. To serve an edge user all the neighboring clusters need
to provide a certain number of degrees of freedom, which leaves
fewer degrees of freedom to serve their own cluster interior users.

\subsection{Inter-cluster Coordination Distance}
In this section, we present one method for grouping the users into
cluster interior and cluster edge users, which will be employed in
our simulations to illustrate our algorithms' performance. Our
proposed metric is based on the channel model in this paper, which
includes Rayleigh fading, shadowing and path loss, and
omnidirectional antennas. With this model, users near the cluster
edge will have low signal power and high interference on average,
and require inter-cluster coordination. Therefore, we do user
grouping based on user locations, and determine an
\emph{inter-cluster coordination area} by the \emph{coordination
distance}, which is defined as follows and illustrated in Fig.
\ref{figICCdiagram}.
\begin{mdef}
\emph{Coordination distance}, $D_c$, is the boundary between
interior and edge users. If the distance of the user to the cluster
edge is no larger than $D_c$, this user is classified as a
\emph{cluster edge user}; otherwise, it is a \emph{cluster interior
user}.
\end{mdef}
In a real implementation, this grouping could be performed based on average signal strength measurements (as employed in the handoff algorithm for example). We defer development of measurement based approaches, however, to future work.

Naturally there is a tradeoff when choosing $D_c$. If $D_c$ is large, more users will be
treated as edge users and enjoy a substantial interference
reduction, but the total throughout will be reduced as the total
number of active users will be reduced. To balance fairness to edge users and the total sum rate, we will investigate
the \emph{mean minimum rate} and \emph{effective sum rate}, as a function of $D_c$.

\paragraph*{Mean Minimum Rate} Suppose that the mobile users are randomly distributed
within each cluster. For a given $D_c$, for each realization of user
locations, denote $R_{min}(D_c)$ as the minimum rate among all the
users in the cluster. \emph{Mean minimum rate}\footnote{Other
similar performance metrics regarding the fairness to the edge users
can also be applied, e.g., the achievable rate at a certain outage
probability. The results, however, will not change.},
$\bar{R}_{min}(D_c)$, is the mean value of $R_{min}(D_c)$, which is
mainly determined by the edge users and will increase as $D_c$
increases.

\paragraph*{Effective sum rate} As the edge user is served by
multiple neighboring clusters, effectively its rate is shared by
those clusters. If there are $N_{c,i}$ clusters serving user $i$,
which is decided by its location and $D_c$, then the effective rate
of this user for each coordinating cluster is $R_i/N_{c,i}$, where
$R_i$ is given as follows according to \eqref{equ3_4}
\begin{equation}
R_i=\frac{1}{B}{\log_2{\Big|\mathbf{I}_{N_r}+{\hat{\mathbf{H}}}_i^{(c)}{\mathbf{T}}_i^{(c)}\mathbf{Q}_i^{(c)}{\mathbf{T}}_i^{(c)*}{\hat{\mathbf{H}}}_i^{(c)*}\Big|}}.
\end{equation}
The \emph{effective sum rate} for each cluster is defined as
\begin{equation}
R_{sum}(D_c)=\sum_{k=1}^K\frac{R_k}{N_{c,k}(D_c)},
\end{equation}
which will decrease with the increase of $D_c$ as more users become
edge users.

For a home cluster, if all the users are interior users, then the
effective sum rate is the conventional sum rate for this home
cluster; if there is an edge user in this home cluster served by
$N_c$ clusters, only $1/N_c$ of this user's rate is counted into the
effective sum rate of each serving cluster, including the home
cluster. Therefore, the effective sum rate is the same for each
cluster in a homogeneous network.

According to these definitions, $\bar{R}_{min}(D_c)$ and
$R_{sum}(D_c)$ characterize the opposing objectives of fairness to
edge users and total sum throughput. We propose to use a utility
function, $U(D_c)$, to evaluate the effect of $D_c$ on both
$\bar{R}_{min}$ and $R_{sum}$.
\begin{mdef}[Utility Function $U(D_c)$]
The utility function $U(D_c)$ is defined by
\begin{equation}
U(D_c)=\alpha\frac{\bar{R}_{min}(D_c)}{\max_{D_c}{\bar{R}_{min}(D_c)}}+(1-\alpha)\frac{R_{sum}(D_c)}{\max_{D_c}{R_{sum}(D_c)}},0\leq\alpha\leq{1}.
\end{equation}
where $\alpha$ is a variable reflecting the design objective. If it
is more valuable to care about edge users, we can pick
$\alpha\rightarrow{1}$; if sum rate is more important, we can pick
$\alpha\rightarrow{0}$. As an example, we pick $\alpha=1/2$, which
means we treat relative changes of $\bar{R}_{min}$ and $R_{sum}$ as
of equal value to the system.
\end{mdef}

\begin{figure}[htb]
\centering
\includegraphics[width=4in]{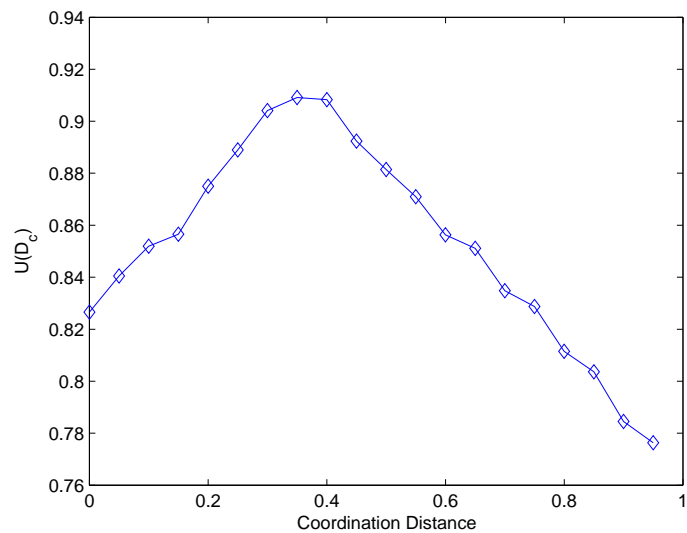}
\caption{$U(D_c)$ for different $D_c$, $R=1$ km.}\label{fig_VD}
\end{figure}

Simulation results of $U(D_c)$ for $D_c\in[0, R)$ \footnote{When
$D_c=R$, the area around the BTSs is classified as inter-cluster
coordination area, which is not going to be the case as the nearby
BTS can provide a high SINR for the users in this area. Therefore,
we only consider $D_c\in[0, R)$ in this simulation.} are shown in
Fig. \ref{fig_VD}, with $B=3$, $R=1$ km and $K=30$, and
interference-free SNR at the cell edge is $18$ dB. Totally $1000$
realizations of user locations are run, and for each realization
$1000$ iterations are simulated with independent channel state. PF
scheduling is used to provide fairness, and the scaled water-filling
power allocation is used for computational efficiency. From the
results we can see that the maximum value is achieved around
$D_c=0.35R$, which will be a proper choice.

Inter-cluster coordination may be designed for criterions other than
$\bar{R}_{min}$ and $R_{sum}$, but the idea of making a good
tradeoff between the fairness for the edge users and the sum rate
persists.

\subsection{Cluster Size}
With a fixed $D_c$, if the cluster size is small, the relative
coordination area is large and there will be too many cluster edge
users which will consume lots of the degrees of freedom and lower
the effective sum rate. Alternatively, a large cluster size will
have a relatively small coordination area, which has small sum rate
loss. However, the requirement of full CSI and synchronization will
prohibit a very large cluster size, and due to path loss the users
benefit little from those BTSs far away. Therefore, to select a
suitable cluster size is important for practical systems, which is
also the motivation to propose the clustered coordination.

\subsubsection{Sum Rates for Different Cluster Sizes}
\begin{figure}[htb]
\centering
\includegraphics[width=4in]{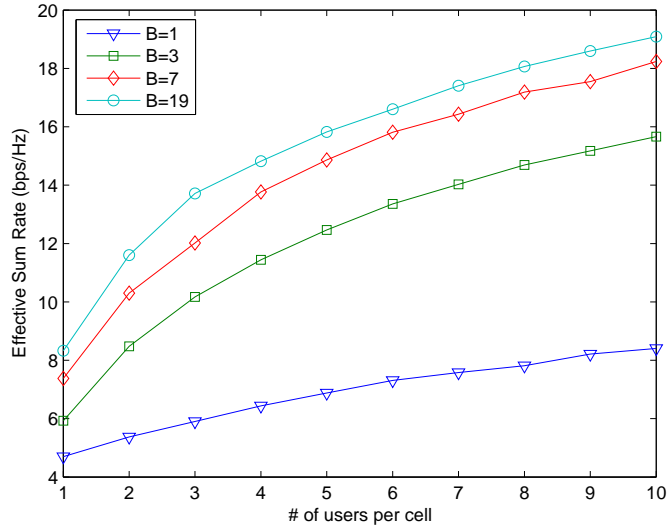}
\caption{Effective sum rate per cell with different cluster size,
$B=1,3,7,19$, $R=1$ km, and $D_c=0.35R$. The standard deviation of
shadowing is $8$dB, the path loss exponent is
$3.7$.}\label{fig_DiffB}
\end{figure}

Fig. \ref{fig_DiffB} shows the effective sum rates per cell for
different cluster sizes, $B= 1, 3, 7, 19$, $D_c=0.35R$, $R=1$ km,
and interference-free SNR at the cell edge is $18$ dB. We can see
that there is a diminishing gain with the increase of the cluster
size: the $3$-cell cluster has a much higher sum rate than the
$1$-cell cluster, and a $7$-cell cluster has a rate gain about $2.5$
bps/Hz over a $3$-cell cluster, while from $B=7$ to $B=19$ the sum
rate increases about $1$ bps/Hz. The lower sum rate for $B=1$ is due
to its relative large edge area. Therefore, a $7$-cell cluster can
already achieve a significant part of the performance gain of the
clustered coordination.

\subsubsection{CSI Feedback Reduction}
The CSI requirement for clustered coordination is on a cluster
scale, which is greatly reduced compared to global coordination.
With $C$ clusters and $B$ cells each cluster, totally there are $BC$
BTSs in the network. For global coordination, the effective channel
matrix for each user is $N_r\times{BCN_t}$, while for clustered
coordination it is $N_r\times{BN_t}$. Therefore, we get the
following lemma:
\begin{lemma}[CSI Reduction]
\emph{For a cellular network with $N_{cell}$ cells, the amount of
CSI feedback for clustered coordination with cluster size $B$ is
$\frac{B}{N_{cell}}$ of that for global coordination.}
\end{lemma}
The amount of CSI feedback for a $7$-cell cluster system is only
$\frac{7}{19}$ of that for a $19$-cell cluster, while the
performance of the $7$-cell cluster system does not degrade much as
shown in Fig. \ref{fig_DiffB}, so a cluster size of $7$ is a
reasonable choice for clustered coordination with the given transmit
power.

\section{Numerical Results}
In this section, the performance of the proposed coordination
strategy is shown via monte carlo simulation. We choose the number
of antennas to be $N_t=4$ and $N_r=2$. The standard deviation of
shadowing is $8$dB, the path loss exponent is $3.7$, and the cell
radius is $1$ km. Other than stated, the interference-free SNR at
the cell edge is $18$ dB, accounting for path loss and ignoring
shadowing and Raleigh fading. We assume all the BTSs in other
clusters transmit at full power. Mobile users are uniformly
distributed within each cluster, and they are associated with
clusters based on locations.

\subsection{Sum Rates for Different Systems}
First, we consider sum rates for different systems with maximum sum
rate (MSR) scheduling. Besides the proposed multi-cell BD systems,
we also compare with the following systems.
\begin{itemize}
\item \emph{Multi-cell DPC with Total Power Constraint (TPC):} This is
an upper bound for the downlink channel of multi-cell systems. We
assume a total power constraint. DPC is applied across BTSs within
the same cluster, and algorithm $2$ in \cite{JinRhe05IT} is used for
power allocation.
\item \emph{Multi-cell BD with TPC:} This is similar to the single-cell
BD, and the water-filling algorithm can be applied to the aggregated
channel for power allocation. This serves as an upper bound for
multi-cell BD with PBPC, and can indicate the capacity loss due to
the PBPC.
\item \emph{TDMA with Intercell Scheduling \cite{Choi06TDMA}:}
Neighboring BTSs cooperatively schedule their transmission, and only
one BTS is active to serve one user at each time slot.
\item \emph{Intercell Scheduling with BD:} Compared to TDMA with
intercell scheduling only, this technique allows one BTS to serve
multiple users at each time slot with BD.
\end{itemize}

\begin{figure*}
\centerline{\subfigure[Different
K]{\includegraphics[width=3.5in]{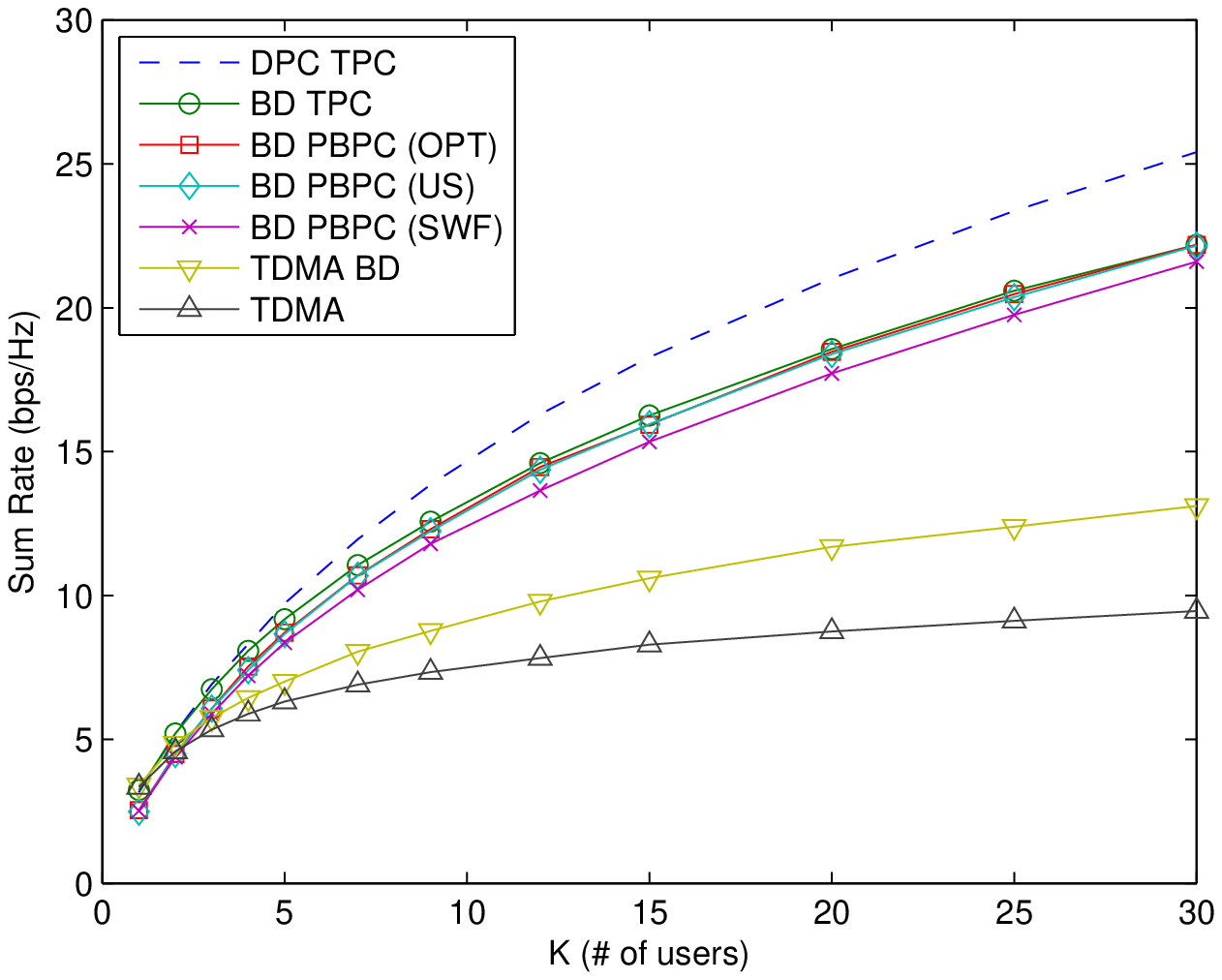} \label{MUBD3B}} \hfil
\subfigure[Small K]{\includegraphics[width=3.5in]{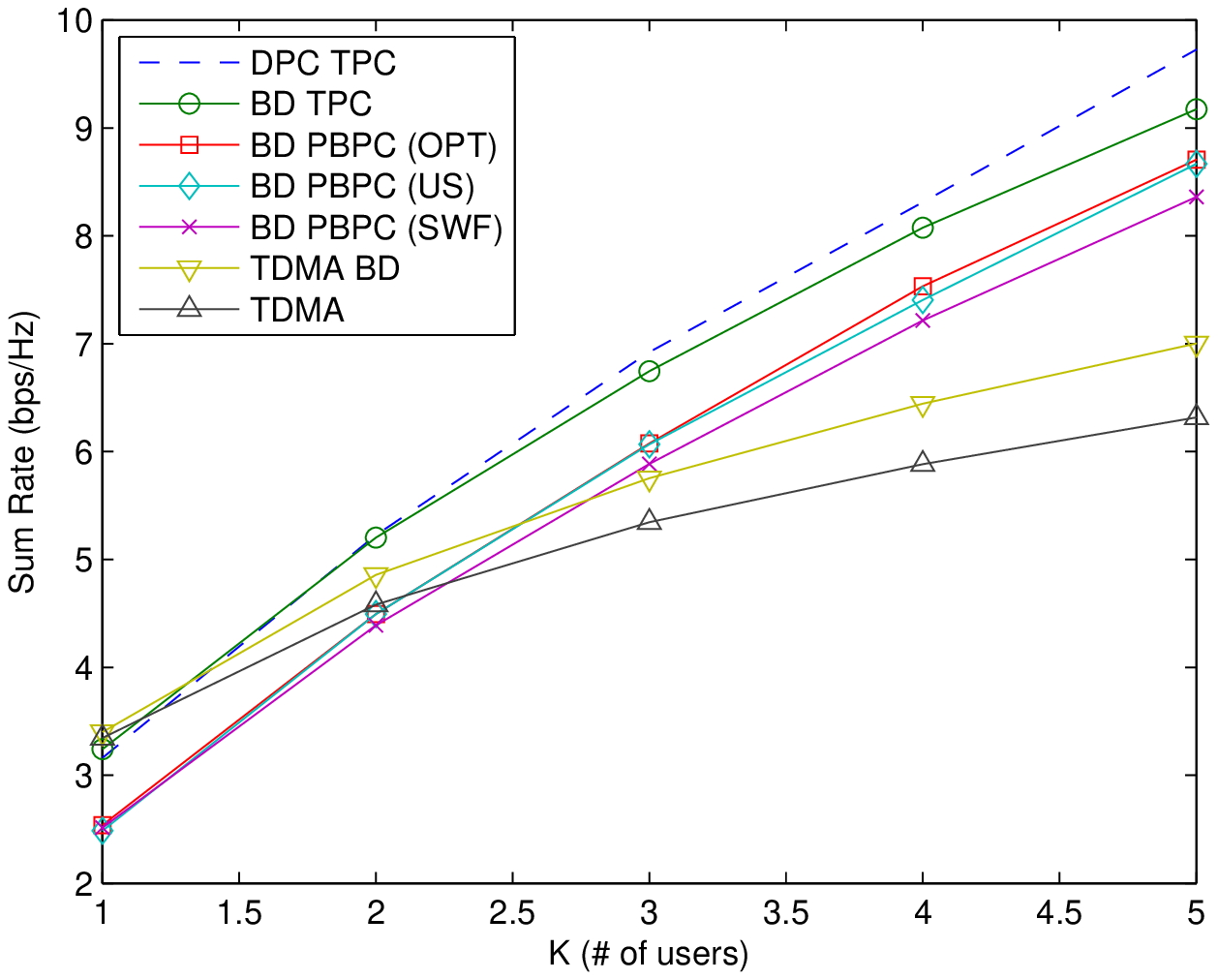}
\label{MUBD3B5U}}} \caption{Sum rate per cell for different systems,
with cluster size $B=3$. ``OPT'' denotes the optimal power
allocation scheme, ``US'' denotes the user scaling scheme, and
``SWF'' denotes the scaled water-filling scheme. ``DPC TPC'' is the
multi-cell dirty paper coding with total power constraint, and
``TDMA'' is the opportunistic intercell scheduling.} \label{fig5_1}
\end{figure*}

Fig. \ref{fig5_1} compares sum rates for different systems. There
are several key observations.
\begin{enumerate}
\item The sum rates of multi-cell BD systems are much higher than
that of the TDMA system with intercell scheduling, and are pretty
close to that of DPC.
\item All multi-cell BD schemes have about the same performance.
\item There is only a marginal rate loss of PBPC to TPC.
\end{enumerate}

\subsection{Distribution of User Rates}
\begin{figure}[htb]
\centering
\includegraphics[width=4in]{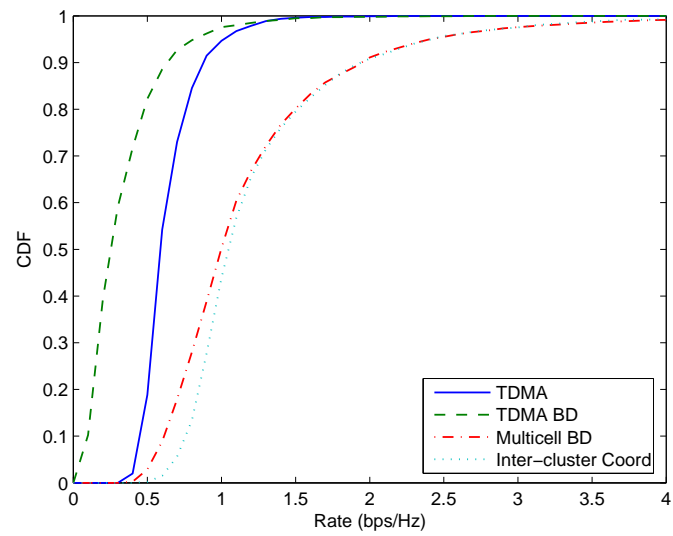}
\caption{CDF of the rates for users in the cluster, $B=3$,
$D_c=0.35R$.}\label{fig_Outage}
\end{figure}

Fig. \ref{fig_Outage} shows the cumulative distribution function
(CDF) of mean rates for users. There are $30$ users uniformly
distributed in the cluster, $B=3$ and $D_c=0.35R$, and PF scheduling
is applied. The simulation setting is similar as that for Fig.
\ref{fig_VD}. We run $100$ realizations for user locations, and for
each realization $1000$ iterations are simulated with independent
channel state and the mean rates are stored. Totally, there are
$3000$ samples of user rates, with which we can plot the CDF. For
example, the rate with $10\%$ outage for intercell scheduling is
$0.4$ bps/Hz, for intercell scheduling with BD is $0.1$ bps/Hz, and
for clustered multi-cell BD with and without inter-cluster
coordination are $0.6$ and $0.8$ bps/Hz, respectively. For
multi-cell BD with inter-cluster coordination, nearly $60\%$ users
have mean rate larger than $1$ bps/Hz and $10\%$ users have mean
rate larger than $2$ bps/Hz, while for intercell scheduling only
less than $5\%$ of users can have mean rate larger than $1$ bps/Hz.

\subsection{Imperfect Channel Knowledge}
\begin{figure}[htb]
\centering
\includegraphics[width=4in]{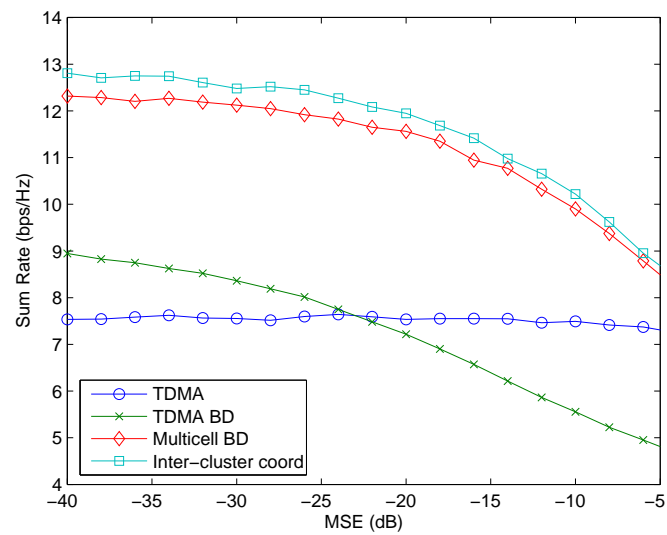}
\caption{Sum rates for different systems with imperfect channel
knowledge.}\label{fig_CSIError}
\end{figure}

Pilot symbols are required for channel estimation, and such training
overhead becomes greater for a larger cluster size. However, there
will be inevitable estimation errors, and the simulation results
accounting for imperfect channel knowledge are shown in Fig.
\ref{fig_CSIError}. The channel estimation model in
\cite{CheAnd04ICC} is used. At the BTSs, the available knowledge of
the small-scale fading channel matrix of the $k$th user is given by
$\check{\mathbf{H}}_k^{(c,b)}=\mathbf{H}_k^{(c,b)}+\mathbf{E}_k^{(c,b)}$,
where $\mathbf{H}_k^{(c,b)}$ is the true channel matrix and
$\mathbf{E}_k^{(c,b)}$ is the channel error. Entries of
$\mathbf{E}_k^{(c,b)}$ follows i.i.d. complex Gaussian distribution
with zero mean and covariance $\sigma^2_{MSE}/2$ per real dimension.
The channel knowledge error is denoted as
MSE$=10\log_{10}\sigma^2_{MSE}$ dB. To demonstrate the impact of
imperfect CSI, we assume equal MSE for each user. The unequal MSE
case is left to future work. We can see that the sum rates for BD
systems decrease as MSE increases, while TDMA system with intercell
scheduling is not sensitive to channel error, but the sum rates of
multicell BD systems are always higher for the simulated range. This
is due to the imperfect inter-user interference cancelation with
channel error for MU-MIMO systems, and such interference is from the
same propagation channel as the information signal, so it will
greatly degrade the performance. Therefore, robust precoding schemes
are required in practical systems.

\section{Conclusion}
In this paper, a clustered BTS coordination strategy is proposed to
increase the available spatial degrees of freedom for MIMO networks,
and thus to reduce interference and increase the sum rate. A cluster
structure is formed, and the users are grouped into cluster interior
users and cluster edge users, served with different coordination
strategies. Cluster interior users are served with intra-cluster
coordination, i.e. multi-cell BD, while cluster edge users are
served by multiple neighboring clusters to reduce inter-cluster
interference. The precoder for multi-cell BD and system parameters
for inter-cluster coordination are designed. It is shown that a
small cluster size (such as $7$) is enough to provide the benefits
of the clustered coordination while greatly reducing the amount of
channel feedback. Numerical results show that the proposed
coordination strategy can provide robust sum rate and edge user rate
gains.

There are many practical issues associated with clustered BTS
coordination, requiring much future work. Compared with global
coordination, the cluster structure reduces the amount of CSI
required at the BTS, but with multiple antennas at both the BTS and
mobiles, the amount of CSI is still daunting. Current schemes are
sensitive to synchronization and CSI error, which is expected to
increase in a cluster system, so robust precoding schemes are
needed. In this paper, we have assumed that all users have perfect
knowledge about other-cluster interference. The investigation of the
imperfect interference estimation is of practical importance and is
a worthy topic of future work. Generally, the analysis of cellular
MIMO networks is an open problem, given the randomness of user
locations, path loss, and matrix channels with fading and shadowing.

\bibliographystyle{IEEEtran}
\bibliography{BD}

\begin{table}[ht]
\caption{System Parameters} \centering
\begin{tabular}{c|c}
\hline \hline Symbol & Description\\
\hline \hline $P$ & the maximum transmit power at each BTS\\
\hline B & number of BTSs in each cluster\\
\hline C & number of clusters we consider\\
\hline K & number of users per cluster\\
\hline $l_k$ & length of data symbol for user $k$\\
\hline $N_t$ & number of transmit antennas at each BTS\\
\hline $N_r$ & number of receive antennas at each mobile\\
\hline $R$ & radius of each cell\\
\hline $D_c$ & coordination distance\\
 \hline\hline
\end{tabular}
\end{table}

\begin{table}[ht]
\caption{User Selection Algorithm} \centering
\begin{enumerate}
\item{Initially, set $\mathcal{S}=\emptyset$ and
$\mathcal{U}=\{1,2,\cdots,K\}$. Set $C_{old}=0$.}
\item While $|\mathcal{S}|<K$ and
$|\mathcal{S}|<\frac{B{N_t}}{N_r}$
\begin{enumerate}
\item for every $k\in\mathcal{U}$
\begin{enumerate}
\item $\hat{\mathcal{S}}=\mathcal{S}+\{k\}$.
\item Calculate $C_{new}=\sum_{s\in\hat{\mathcal{S}}}f_s$.
\item if $C_{new}>C_{old}$, set $C_{old}=C_{new}$, and $\hat{k}=k$.
\end{enumerate}
\item Let $\mathcal{S}=\mathcal{S}+\{\hat{k}\}$,
$\mathcal{U}=\mathcal{U}-\{\hat{k}\}$.
\end{enumerate}
\end{enumerate}
\end{table}

\end{document}